\documentclass[journal]{IEEEtran}
\usepackage{graphicx} 
\usepackage[style=ieee,sorting=none,giveninits=true]{biblatex}
\graphicspath{ {./figs/} }
\usepackage{caption}
\usepackage{subcaption}
\usepackage{threeparttable}
\usepackage{url}
\usepackage{multirow}
\usepackage{float}
\usepackage{tikz}
\usepackage{amsmath}
\usepackage{amssymb}
\usepackage{amsfonts}
\usepackage{soul}
\usepackage{tikz}
\usetikzlibrary{arrows.meta, positioning, shapes.geometric, decorations.pathreplacing}

\begin{document}

\title{On the Role of Time Series Clustering \\in Traffic Matrix Prediction}

\author{Martha~Cash,
        Charlotte~Fowler,
        and~Alexander M.~Wyglinski}



\maketitle

\begin{abstract}
This paper analyzes the role of time-series clustering in traffic matrix (TM) prediction. Traffic flows within a TM often exhibit heterogeneous behavior, which can reduce the effectiveness of global forecasting models that predict all flows jointly. To address this, we propose a clustering-based prediction framework that groups flows into smaller subsets and trains separate predictors for each group. We evaluate four traffic-flow representations for clustering, namely histogram, autocorrelation function (ACF), power spectral density (PSD), and na\"{i}ve partitioning, and analyze how both the representation choice and the number of clusters affect prediction performance.
Experiments using the publicly available Abilene and G\'EANT datasets show that clustering consistently improves over global forecasting baselines, while remaining substantially less costly than local prediction. The results further show that most of the performance gain is achieved at moderate values of \(K\), with diminishing returns as the number of clusters increases. Although different clustering representations produce different partitions of the traffic flows, they often achieve similar root mean squared error (RMSE). This suggests that the main benefit of clustering lies in decomposing the TM prediction task into smaller subproblems, while the exact cluster structure plays a more limited role in determining overall prediction accuracy.
\end{abstract}


%
\IEEEpeerreviewmaketitle

\section{Introduction}
Traffic matrices (TMs), which quantify demand between source-destination (SD) node pairs in a network, are a central input for traffic engineering (TE) tasks such as routing optimization and network resource allocation~\cite{TMPrimer}. In practice, TMs are difficult to measure accurately and collect in real-time~\cite{soule2005}. As a result, predicting future TMs from historical observations has become an important capability for supporting proactive network management~\cite{azzouni2018neutm}.

Traffic matrix prediction is challenging because TMs are high-dimensional and their  traffic flows often exhibit diverse temporal behavior. Recent work has increasingly relied on deep learning (DL) models for TM prediction~\cite{azzouni2018neutm, aloraifan2021deep, jiang2022internet}, with most approaches adopting a global forecasting paradigm in which a single model predicts all traffic demands simultaneously. Global models can be effective when the underlying features exhibit similar patterns~\cite{MonteroManso2020}, but their accuracy can degrade when the data are heterogeneous~\cite{Bandara2017}. This challenge is especially relevant for TM prediction, where different traffic flows may exhibit distinct temporal patterns. 

\begin{figure}[!t]
    \centering
    \includegraphics[width=0.48\textwidth]{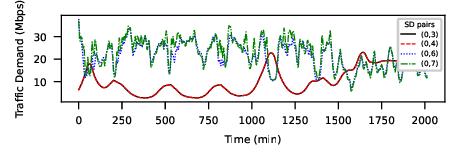}
    \caption{Example traffic flows from different source-destination pairs in the Abilene~\cite{AbileneDataset} dataset, illustrating heterogeneous behavior across flows.}
    \setlength{\belowcaptionskip}{-30pt}
    \label{fig:traffic_flows}
    \vspace{-15pt}
\end{figure}

As illustrated in Fig.~\ref{fig:traffic_flows}, we observe that traffic demands within a TM often differ significantly in behavior. Some SD pairs, such as $(0,3)$ and $(0,4)$, exhibit similar trends, whereas others, such as $(0,4)$ and $(0,7)$, deviate markedly. This heterogeneity has motivated the development of increasingly complex DL architectures, including Attention-based Convolutional Recurrent Neural Networks (ARCNN)~\cite{gao2020incorporating} and Convolutional Long Short-Term Memory Networks (CNN-LSTM)~\cite{Jiang2022}, which are designed to model diverse flow behavior within a TM.

Rather than further increasing the complexity of global TM prediction models, this paper investigates clustering as an alternative way to address flow heterogeneity. The central idea is to group traffic flows with similar behavior and train separate predictors for each group, thereby decomposing the overall TM prediction task into smaller subproblems. In this sense, clustering provides an intermediate strategy between entire-matrix prediction, which uses a single global model, and local prediction, which trains one model per flow.

In the authors' prior work~\cite{cash2025improving}, it was demonstrated that clustering improved TM prediction performance. In this paper, we extend our existing work to investigate which representations of traffic flows are the most useful for clustering. We investigate how different traffic flow representations change prediction accuracy. Additionally, the sensitivity of clustering-based TM prediction to the number of clusters remains unclear; therefore, this work evaluates RMSE across a range of $K$ values to identify effective operating points for each clustering method. Furthermore, we investigate whether the performance gains arise from meaningful clustering of the traffic flows or from simply decomposing the problem itself.

To address these questions, we propose a clustering-based TM prediction framework using one of four possible representations: histogram, autocorrelation function (ACF), power spectral density (PSD), and na\"{i}ve clustering representations. We evaluate how these representations affect prediction accuracy, runtime, and cluster structure, and we analyze how the choice of the number of clusters $K$ influences the tradeoff between predictive performance and computational cost.

The main contributions of this paper are:
\begin{itemize}
\item A clustering-based TM prediction framework that studies multiple traffic-flow representations, including histogram, ACF, PSD, and na\"{i}ve partitioning, within a common prediction pipeline.
\item A systematic analysis of the effect of cluster granularity on prediction accuracy and runtime through $K$-sweep experiments.
\item An empirical study of how clustering structure relates to prediction performance, including pairwise clustering similarity, cluster-size statistics, and per-flow prediction-error consistency.
\item A comparative evaluation against entire-matrix and local baselines on the Abilene and G\'EANT datasets, showing that clustering improves over global prediction while providing a practical alternative to local forecasting.
\end{itemize}

The remainder of the paper is organized as follows. Section~\ref{sec:related_work} reviews related work. Section~\ref{sec:preliminaries} introduces the TM prediction setting and clustering background. Section~\ref{sec:approach} presents the clustering framework and flow representations. Section~\ref{sec:experimental_settings} describes the datasets and experimental setup. Section~\ref{sec:results} reports the experimental results. Section~\ref{sec:discussion} discusses the broader implications of the results. Finally, Section~\ref{sec:conclusion} concludes the paper.

\section{Related Work}\label{sec:related_work}
Internet traffic prediction can be viewed as a time-series forecasting problem, and a broad range of methods have been explored for this purpose, from classical statistical techniques to deep learning (DL) models. Otoshi \textit{et al.} separated traffic into long- and short-term components and applied a seasonal autoregressive integrated moving average model for forecasting~\cite{otoshi2015traffic}. Although this approach can capture recurring behavior, it relies on repeated differencing to impose stationarity, which limits its practical use. Valadarsky \textit{et al.} investigated Feed Forward Neural Networks, Convolutional Neural Networks (CNNs), and nonlinear autoregressive models for traffic prediction~\cite{valadarsky2017learning}; however, these approaches do not explicitly model temporal dependence and therefore yield weaker predictive performance.

To better capture dependencies over time, more recent works~\cite{zhao2018towards, troia2018deep, ramakrishnan2018network} have adopted Recurrent Neural Networks (RNNs), including Long Short-Term Memory (LSTM) and Gated Recurrent Unit (GRU) models, for entire-TM prediction. Other studies~\cite{le2021ai, zheng2022flow, li_network_2024} instead treated each TM entry as a separate time series and train an individual predictor for each flow. Liu \textit{et al.} proposed a hybrid strategy in which total traffic volume is predicted first and then distributed across TM elements using precomputed ratios based on mean traffic volume, with additional correction models for high-volume flows~\cite{liu2014prediction}. Nevertheless, several studies~\cite{liu2019traffic, gao2020incorporating} have shown that predicting TM elements independently can ignore correlations between flows, which may reduce overall prediction accuracy.

TMs can also be viewed as structured collections of related time series. In this broader setting, several works have explored clustered time-series forecasting as a way to handle heterogeneous data. Aghabozorgi \textit{et al.} categorized time-series clustering approaches into distance-based, feature-based, and model-based methods~\cite{aghabozorgi2015clustering}. This distinction is useful in the context of this work because the clustering outcome depends not only on the algorithm itself, but also on how the time series are represented before clustering.

Distance-based clustering operates directly on the raw sequences or transformed sequences and is strongly influenced by the dissimilarity measure used. Effective distances must account for noise, unequal scales, and other variations across time series~\cite{Ma2017distance}. However, many shape-based distances can produce groupings that are difficult to interpret in an application setting~\cite{wang2006characteristic}. Model-based clustering instead summarizes each time series through model parameters and compares series indirectly through those parameter sets. Common examples include Gaussian mixture models (GMMs) and hidden Markov models (HMMs)~\cite{paparrizos2024}. These methods can be effective, but they also rely on assumptions about the underlying data-generating process. By contrast, histogram-based density estimates provide a distributional representation of time series without requiring strong modeling assumptions~\cite{arroyo2009forecasting, balzanella2020histogram}.

Feature-based clustering provides another alternative by representing each time series through a selected set of summary features. Bandara \textit{et al.} distinguished between approaches that extract a very large number of features and those that rely on a smaller set of interpretable features~\cite{bandara2020forecasting}. Fulcher \textit{et al.} proposed an automated pipeline that extracts more than 900 features from each series, but such large feature sets are often difficult to interpret and may be impractical in many applications. In contrast, a smaller set of carefully chosen features can capture important aspects of time-series behavior while remaining easier to interpret~\cite{bandara2020forecasting, nanopoulos2001feature}. This perspective is particularly relevant for our study, where we consider representations that emphasize distributional behavior, autocorrelation, and frequency content.

Finally, a range of clustering algorithms can be used once a representation and dissimilarity measure have been selected. Common choices include partitioning methods such as $k$-means and hierarchical clustering~\cite{paparrizos2015kshape, jain2025hierarchical, radovanovic2020HAC}. In this work, hierarchical agglomerative clustering (HAC) is especially appealing because it operates directly on pairwise dissimilarity matrices, making it compatible with the different distance measures used for the histogram, ACF, and PSD representations. In addition, the hierarchical structure supports systematic analysis across different values of $K$, which is central to the present study. Compared with centroid-based methods such as $k$-means, hierarchical clustering is also generally more robust to outliers and does not require Euclidean centroids.

Although clustered forecasting has been studied for heterogeneous time-series data~\cite{radovanovic2020HAC, bandara2020forecasting, cash2025improving}, prior work has not systematically investigated clustering-based prediction for internet TMs. To the best of our knowledge, it remains unclear which traffic-flow representations are most useful for clustering, how sensitive TM prediction is to the number of clusters, and whether the gains from clustering arise from meaningful grouping structure or from decomposition of the prediction task itself. These questions motivate the framework and analysis developed in this paper.

\section{Time Series Clustering for Traffic Matrix Prediction}
\label{sec:preliminaries}
This section formalizes the TM prediction setting, highlights the limitations of existing EM and local prediction approaches, and introduces cluster-based prediction as an intermediate framework. We also summarize the elements of whole time-series clustering needed to motivate the representation and clustering choices studied in this work.

\subsection{Traffic Matrix Prediction Overview}
For a network with $N$ nodes, a TM is an $N \times N$ matrix where each entry $(i,j)$ represents the traffic volume from source node $i$ to destination node $j$, typically in bytes. Let $\mathrm{TM}_t\in \mathbb{R}^{N\times N}$ denote the TM observed at discrete time index $t$, and is given by: 

\begin{equation}
    \label{eq:traffic_matrix}
    \mathrm{TM_t} = \begin{bmatrix}
    T_{1,1}^t & T_{1,2}^t & \cdots & T_{1,N}^t \\
    T_{2,1}^t & T_{2,2}^t & \cdots & T_{2,N}^t \\
    \vdots    & \vdots    & \ddots & \vdots    \\
    T_{N,1}^t & T_{N,2}^t & \cdots & T_{N,N}^t \\
    \end{bmatrix}
\end{equation}

Over a sequence of observations $t=1, \ldots, T$ each entry $(i,j)$ forms a univariate traffic flow time series. Each $(i,j)$ entry may be referred to as a traffic flow or origin-destination (OD) pair. We use traffic flow and OD pair interchangeably throughout this work. This perspective allows TM prediction to be viewed either as forecasting a multivariate object jointly or as forecasting its constituent traffic flows individually.

The goal of TM prediction is to forecast $\mathrm{TM}_{t+1}$ given a history of $k$ previous TMs: $(\mathrm{TM}_{t-k}, \dots, \mathrm{TM}_t)$. TMs exhibit nonlinear, nonstationary, and periodic behavior, making prediction challenging. DL models are well-suited for this task due to their ability to capture such complex patterns. We formulate prediction as a supervised learning problem, following the approach in~\cite{valadarsky2017learning}.

\subsection{Limitations of Existing Approaches}
TM prediction approaches generally fall into two categories: entire-matrix (EM) prediction and local prediction~\cite{liu2014prediction, zheng2022flow, azzouni2018neutm}. EM prediction uses a single model to forecast the whole TM, while local prediction trains one model per flow, resulting in $N^2$ models for a network with $N$ nodes. Although EM prediction is computationally efficient, its accuracy can be limited by the need to model heterogeneous traffic flows within a single global predictor. In contrast, local prediction often yields higher accuracy~\cite{zheng2022flow}, but it is computationally expensive and scales poorly as the number of flows increases.

A key limitation of EM prediction is its implicit assumption of homogeneity among flows, an assumption that rarely holds in practice. Gao \textit{et al}.~\cite{gao2020incorporating} showed that flows from the same source are often correlated, while flows from different sources may not be. This indicates that TMs contain latent structure that can be exploited for prediction. This observation motivates the present work. Cluster-based prediction occupies a middle ground between EM and local modeling: flows assigned to the same cluster share a predictor, allowing the model to exploit common structure without forcing all flows into a single global representation. We therefore investigate whether time-series clustering can improve TM prediction accuracy relative to EM prediction while reducing the computational burden relative to fully local models.

\subsection{Clustering Based Prediction Framework}

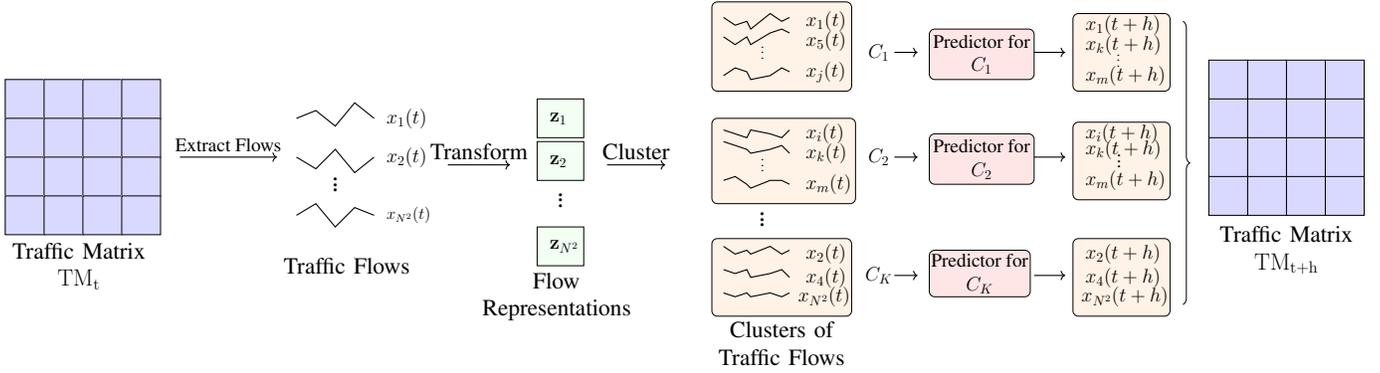
\begin{figure*}[t]
\centering
\resizebox{\linewidth}{!}{%
\begin{tikzpicture}
    
    \foreach \i in {0,1,2,3} {
        \foreach \j in {0,1,2,3} {
            \draw[fill = blue!15] (\j,-\i) rectangle ++(1,-1);      
        }
    }
    
    \node[font=\LARGE, align=center] at (1.9,-4.8) {Traffic Matrix\\$\mathrm{TM_t}$};
    
    \draw[->, thick] (4.5,-2) -- (7,-2) node[midway, above, font=\Large] {Extract Flows};
    
    \draw[thick] (7.5,-1) -- (8,-0.8) -- (8.5,-1.2) -- (9,-0.6) -- (9.5,-1);
    \node[right, font=\Large] at (9.7,-1) {$x_1(t)$};
    
    \draw[thick] (7.5,-2) -- (8,-2.3) -- (8.5,-1.8) -- (9,-2.4) -- (9.5,-2);
    \node[right, font=\Large] at (9.7,-2) {$x_2(t)$};
    
    \node[font=\huge] at (8.5,-2.6) {$\vdots$};
    
    \draw[thick] (7.5,-3.5) -- (8,-3.2) -- (8.5,-3.8) -- (9,-3.3) -- (9.5,-3.5);
    \node[right, font=\large] at (9.7,-3.5) {$x_{N^2}(t)$};
    
    \node[font=\LARGE,align = center] at (8.8,-4.8) {Traffic Flows};
    
    \draw[->, thick] (11.4,-2.2) -- (13,-2.2) node[midway, above, font=\LARGE, align=center] {Transform};
    
    
    \draw[fill=green!5] (13.7,-0.5) rectangle ++(1.2,-1); 
    \node[right, font=\Large] at (13.9,-1) {$\mathbf{z}_1$};
    
    \draw[fill=green!5] (13.7,-1.6) rectangle ++(1.2,-1);
    \node[right, font=\Large] at (13.9,-2.1) {$\mathbf{z}_2$};
    
    \node[font=\huge] at (14.3,-3) {$\vdots$};
    
    \draw[fill=green!5] (13.7,-3.8) rectangle ++(1.2,-1);
    \node[right, font=\Large] at (13.9,-4.3) {$\mathbf{z}_{N^2}$};
    \node[font=\LARGE, align=center] at (14.2,-5.6) {Flow\\Representations};
    
    \draw[->, thick] (15.5,-2.2) -- (17,-2.2) node[midway, above, font=\LARGE] {Cluster};
    \draw[rounded corners, fill=orange!10] (18.2,2) rectangle ++(3.6,-2.3);
    \node[font=\Large] at (22.5,0.7) {$C_1$};
    
    \draw[thick] (18.5,1.6) -- (18.8, 1.4) -- (19.1, 1.5) -- (19.2,1.3) -- (19.7,1.7) -- (20.0,1.5) -- (20.2,1.6);
    \node[right, font=\Large] at (20.5,1.5) {$x_1(t)$};
    
    \draw[thick] (18.5,1.1) -- (18.8,0.9) -- (19.1,1.1) -- (19.2,0.9) -- (19.7,1.2) -- (20.0,1.3) -- (20.2,1.2);
    \node[right, font=\Large] at (20.5, 1) {$x_5(t)$};
    
    \node at (19.5, 0.8) {$\vdots$};
    
    \draw[thick] (18.5, 0.1) -- (18.8,0.3) -- (19.1, 0.2) -- (19.2,0) -- (19.7,0.1) -- (20.0,0.3) -- (20.2,0.2);
    \node[right, font=\Large] at (20.5, 0.2) {$x_j(t)$};

    \draw[->, thick] (22.9, 0.7) -- (23.5, 0.7);
   \draw[rounded corners, fill=red!10] (23.8, 1.3) rectangle ++(2.7,-1.3)  node[midway, align = center, font=\Large] {Predictor for \\$C_1$};
   \draw[->, thick] (26.5, 0.7) -- (27.3, 0.7);

    \draw[rounded corners, fill=orange!10] (27.5,1.7) rectangle ++(2.5,-2.0) ;
    \node[right, font=\Large] at (27.7, 1.4) {$x_1(t+h)$};
    \node[right, font=\Large] at (27.7, 0.9) {$x_k(t+h)$};
    \node[right] at (28.5, 0.6) {$\vdots$};
    \node[right, font=\Large] at (27.7, 0.1) {$x_m(t+h)$};

    \draw[rounded corners, fill=orange!10] (18.2,-1.0) rectangle ++(3.7,-2.2);
    \node[font=\Large] at (22.5,-2.0) {$C_2$};
    
    \draw[thick] (18.5, -1.3) -- (18.8,-1.4) -- (19.1, -1.5) -- (19.2, -1.3) -- (19.7,-1.4) -- (20.0,-1.5) -- (20.2,-1.3);
    \node[right, font=\Large] at (20.5, -1.4) {$x_i(t)$};
    
    \draw[thick] (18.5, -1.6) -- (18.8,-1.7) -- (19.1, -1.8) -- (19.2, -1.9) -- (19.7,-1.8) -- (20.0,-1.7) -- (20.2,-1.9);
    \node[right, font=\Large] at (20.5, -1.9) {$x_k(t)$};
    
    \node at (19.5, -2.1) {$\vdots$};
    
    \draw[thick] (18.5, -2.6) -- (18.8,-2.5) -- (19.1, -2.7) -- (19.2, -2.8) -- (19.7,-2.6) -- (20.0,-2.6) -- (20.2,-2.7);
    \node[right, font=\Large] at (20.5, -2.7) {$x_m(t)$};

    \draw[->, thick] (22.9, -2.0) -- (23.5, -2.0);
    \draw[rounded corners, fill=red!10] (23.8,-1.4) rectangle ++(2.7,-1.3)  node[midway, align=center, font=\Large] {Predictor for\\$C_2$};
    \draw[->, thick] (26.5, -2.0) -- (27.3, -2.0);

    \draw[rounded corners, fill=orange!10] (27.5, -1.1) rectangle ++(2.5,-2.0) ;
    \node[right, font=\Large] at (27.7, -1.4) {$x_i(t+h)$};
    \node[right, font=\Large] at (27.7, -1.8) {$x_k(t+h)$};
    \node[right, font=\Large] at (28.5, -2.0) {$\vdots$};
    \node[right, font=\Large] at (27.7, -2.6) {$x_m(t+h)$};

    \node[font=\huge] at (19.5, -3.5) {$\vdots$};
    \draw[rounded corners, fill=orange!10] (18.2,-4.1) rectangle ++(3.6,-2.0);
    \node[font=\Large] at (22.5,-5.05) {$C_K$};
    
    \draw[thick] (18.5, -4.3) -- (18.8,-4.5) -- (19.1, -4.4) -- (19.2, -4.5) -- (19.7,-4.3) -- (20.0,-4.5) -- (20.2,-4.4);
    \node[right, font=\Large] at (20.5, -4.5) {$x_2(t)$};
    
    \draw[thick] (18.5, -4.9) -- (18.8,-5.0) -- (19.1, -4.9) -- (19.2, -5.1) -- (19.7,-5.0) -- (20.0,-4.9) -- (20.2,-5.1);
    \node[right, font=\Large] at (20.5, -5.1) {$x_4(t)$};
    
    \draw[thick] (18.5, -5.5) -- (18.8,-5.6) -- (19.1, -5.5) -- (19.2, -5.6) -- (19.7,-5.5) -- (20.0,-5.6) -- (20.2,-5.5);
    \node[right, font=\Large] at (20.3, -5.6) {$x_{N^2}(t)$};

    \draw[->, thick] (22.9, -5.05) -- (23.5, -5.05);

    \draw[rounded corners, fill=red!10] (23.8,-4.4) rectangle ++(2.6,-1.2)  node[midway, align=center, font=\Large] {Predictor for\\$C_K$};

    \draw[->, thick] (26.5, -5.05) -- (27.3, -5.05);
   
    \draw[rounded corners, fill=orange!10] (27.5,-4.1) rectangle ++(2.5,-2.0) ;
    
    \node[right, font=\Large] at (27.7, -4.5) {$x_2(t+h)$};
    
    \node[right, font=\Large] at (27.7, -5.1) {$x_4(t+h)$};
    
    \node[right, font=\Large] at (27.6, -5.6) {$x_{N^2}(t+h)$};

    \node[font=\LARGE, align=center] at (20.0,-6.8) {Clusters of \\Traffic Flows}; 
    \draw [decorate, decoration={brace, amplitude=5pt, raise=4ex}]
    (29.7,1.5) -- (29.7,-5.8);

    \foreach \i in {0,1,2,3} {
        \foreach \j in {0,1,2,3} {
            \draw[fill = blue!15] (\j+31,\i-2.5) rectangle ++(1,-1);      
        }
    }
    \node[font=\LARGE, align=center] at (33,-4.4) {Traffic Matrix\\$\mathrm{TM_{t+h}}$};

\end{tikzpicture}
}

\caption{Overview of the proposed cluster-based traffic matrix prediction framework. Individual traffic flows are extracted from the traffic matrix time series, transformed into feature representations, clustered using hierarchical clustering, and used to train cluster-specific prediction models. The resulting flow predictions are then aggregated to reconstruct the predicted traffic matrix. }
\label{fig:clustering_framework}
\end{figure*}

As illustrated in Fig.~\ref{fig:clustering_framework}, the proposed framework first extracts individual traffic flows from a TM, transforms each flow into a representation suitable for clustering, groups similar flows into clusters, and then trains one predictor per cluster. The cluster-specific forecasts are aggregated to reconstruct the predicted TM, namely, $\mathrm{TM}_{t+1}$. More generally, the framework supports prediction over a horizon $h$. In this work, we focus on one step ahead prediction, $h=1$. 

Let $\mathcal{X} =\{x_1,\ldots, x_M \}$ denote the set of traffic flows extracted from the TM sequence, where $x_i \in \mathbb{R}^{T}$ is a univariate series of length $T$ and $M = N \times N$. In this work, clustering is performed over these whole traffic flows. Clustering is an unsupervised learning approach to classifying data when there is no knowledge about the ground-truth classes~\cite{paparrizos2024}. Generally, the goal of clustering is to partition $\mathcal{X}$ into $K$ clusters $\mathcal{C} = \{c_1, c_2, \ldots, c_K \}, \,\, K \leq M$. Then, $\mathcal{X} = \cup_{i=1}^{K} \mathcal{C}_i$ and $\mathcal{C}_i \cap \mathcal{C}_j = \varnothing, i\neq j$~\cite{aghabozorgi2015clustering}. We consider whole time-series clustering, where  given a time series dataset $\mathcal{X}$, the individual time series $x_i$ are clustered into distinct groups, and every $x_i$ contributes to determining the intra and inter-clustering relationships across $\mathcal{X}$~\cite{aghabozorgi2015clustering}. 

The pipeline for whole time series clustering has three main parts: representation, dissimilarity measure, and clustering procedure~\cite{paparrizos2024}. These components correspond to the transform and cluster stages of Fig.~\ref{fig:clustering_framework}. First, a representation maps each raw flow into a feature space that emphasizes specific temporal characteristics. Then, a dissimilarity measure is used to quantify pairwise differences between represented flows. Common dissimilarity measures include euclidean distance, dynamic time warping, and Jaccard similarity~\cite{paparrizos2024}. Finally, the clustering algorithm groups flows according to the pairwise distance relationship. We consider HAC for the clustering procedure in this work~\cite{patel2015HAC}. HAC is an approach for grouping objects into clusters through a hierarchical sequence of merges. Agglomerative clustering begins with each traffic flow in its own singleton cluster and iteratively merges until $K$ clusters are formed or all time series form one cluster. The merging criterion between iterations is based on the linkage criterion between clusters~\cite{Li2017impact}. 

The following section discusses the different representations and distance measures considered in the framework. The goal of this work is to understand how representation of the traffic flows in TMs affects prediction performance. Additionally, because hierarchical clustering does not by itself determine an optimal number of clusters, we also study how prediction performance varies with $K$. 

\section{Proposed Time Series Representation and Clustering Methods}
\label{sec:approach}
This section describes the traffic-flow representations used to evaluate the proposed clustering-based prediction framework. By studying multiple representations, we aim to understand how emphasizing different properties of traffic flows affects the induced clustering structure and, in turn, prediction performance. Specifically, we consider three representations that capture complementary aspects of flow behavior: distributional structure, temporal dependence, and frequency-domain content.

First, we consider a histogram-based representation, extending our prior work in~\cite{cash2025improving}. This representation groups flows according to similarities in their empirical value distributions. Second, we study an ACF-based representation~\cite{albino2024time}, which characterizes flows according to their autocorrelation structure and captures similarities in lag-dependent temporal behavior. Third, we consider a power spectral density (PSD) representation, which emphasizes the frequency-domain content of each flow and enables grouping of series with similar dominant periodicities~\cite{rivera2017robust}.

Finally, we propose a na\"{i}ve clustering baseline in which flows are assigned randomly to $K$ clusters without using any feature representation, distance metric, or similarity measure. This baseline helps isolate the effect of partitioning the prediction task into smaller subproblems, independent of any structure induced by representation-based clustering. Across all methods, the downstream prediction pipeline remains unchanged; only the mechanism used to form flow groups differs.

\subsection{Histogram Representation}
Let $x_i = [x_i(1),x_i(2),\dots,x_i(T)]^\top \in \mathbb{R}^T$ denote the normalized traffic flow time series associated with the $i^{th}$ traffic flow, where $i \in \{1,\dots,N^2\}$. To characterize the distributional behavior of each flow, we represent $x_i$ by a histogram-based empirical probability distribution.

Specifically, let $\mathcal{B}=\{B_1,\dots,B_L\}$ denote a partition of the normalized value range into $L$ bins. The histogram representation of flow $x_i$ is then defined as the vector
\begin{equation}
p_i = [p_i(1),p_i(2),\dots,p_i(L)]^\top,
\end{equation}
where
\begin{equation}
p_i(\ell)=\frac{1}{T}\sum_{t=1}^{T}\mathbf{1}\{x_i(t)\in B_\ell\}, \qquad \ell=1,\dots,L,
\end{equation}
and $\mathbf{1}\{\cdot\}$ denotes the indicator function. By construction, $p_i$ is a discrete probability mass function satisfying
\begin{equation}
p_i(\ell)\geq 0, \qquad \sum_{\ell=1}^{L} p_i(\ell)=1.
\end{equation}

This representation emphasizes the empirical distribution of traffic values rather than their temporal ordering. As a result, flows with similar value distributions, such as burstiness, sparsity, or multimodal behavior, can be grouped together even if they originate from different nodes.

To quantify similarity between two histogram representations $p_i$ and $p_j$, we use the Jensen-Shannon divergence (JSD)~\cite{menendez1997jensen}, which is a symmetric and bounded variant of the Kullback-Leibler divergence. Let
\begin{equation}
m_{ij} = \frac{1}{2}(p_i+p_j)
\end{equation}
denote the midpoint distribution between $p_i$ and $p_j$. The JSD between flows $x_i$ and $x_j$ is defined as
\begin{equation}
\mathrm{JSD}(p_i \,\|\, p_j)
=
\frac{1}{2}D_{\mathrm{KL}}(p_i \,\|\, m_{ij})
+
\frac{1}{2}D_{\mathrm{KL}}(p_j \,\|\, m_{ij}),
\label{eq:JSD}
\end{equation}
where
\begin{equation}\label{eq:KL_div}
D_{\mathrm{KL}}(p \,\|\, q)=\sum_{\ell=1}^{L} p(\ell)\log \frac{p(\ell)}{q(\ell)}
\end{equation}
is the KL divergence. In~\eqref{eq:KL_div}, $D_{\mathrm{KL}}(p \,\|\, q)$ denotes the KL divergence between two discrete probability mass functions defined over the same $L$ histogram bins where $\log$ is taken base 2. For the JSD calculation, $q$ is the midpoint distribution $m_{ij}$, with entries $m_{ij}(\ell)=\frac{1}{2}(p_i(\ell)+p_j(\ell))$ for $\ell=1,\dots,L$. The JSD satisfies $0 \leq \mathrm{JSD}(p_i \,\|\, p_j) \leq 1$, and equals zero if and only if $p_i=p_j$. 

Using \eqref{eq:JSD}, we compute the pairwise dissimilarities between all $N^2$ traffic flows and form the matrix
\begin{equation}
\mathbf{D}^{(\mathrm{hist})} \in \mathbb{R}^{N^2 \times N^2},
\quad
\mathbf{D}^{(\mathrm{hist})}_{ij} = \mathrm{JSD}(p_i \,\|\, p_j).
\end{equation}
This matrix serves as the input to hierarchical agglomerative clustering. We use complete linkage in the HAC stage for the histogram representation. Starting from singleton clusters, HAC iteratively merges the pair of clusters with the smallest inter-cluster dissimilarity. After each merge, the histogram-based distance matrix $\mathbf{D}^{(\mathrm{hist})}$ is updated to reflect the dissimilarity between the newly formed cluster and all remaining clusters.

Under complete linkage, the distance between two clusters $U$ and $V$ is defined as the maximum pairwise distance between their members~\cite{dasgupta2005performance}
\begin{equation}
d_{\mathrm{CL}}(U,V)
=
\max_{x_i \in U,\; x_j \in V}
\mathbf{D}^{(\mathrm{hist})}_{ij}.
\end{equation}

Overall, the histogram-based approach groups traffic flows that are similar in terms of their empirical distributional behavior over time. These clusters provide one representation of flow similarity that is subsequently used to train cluster-specific prediction models within the proposed traffic matrix prediction framework.

\subsection{ACF Representation}

For the ACF-based representation, each traffic flow $x_i \in \mathbb{R}^T$ is characterized by its sample autocorrelation function (ACF). The ACF at lag $\ell$ measures the linear dependence between the flow and a time-shifted version of itself, and is given by
\begin{equation}
\rho_i(\ell) = \mathrm{Corr}\!\bigl(x_i(t),x_i(t-\ell)\bigr),
\qquad \ell = 1,\dots,L,
\label{eq:acf}
\end{equation}
where $\ell$ denotes the lag and $L$ is the maximum lag considered. The choice of $L$ is discussed in Section~\ref{sec:experimental_settings}.

Using \eqref{eq:acf}, we represent the $i^{th}$ traffic flow by its vector of autocorrelation values,
\begin{equation}
\mathbf{f}_i^{(\mathrm{acf})}
=
\begin{bmatrix}
\rho_i(1) & \rho_i(2) & \cdots & \rho_i(L)
\end{bmatrix}^{\top}.
\end{equation}
This representation preserves lag-dependent temporal structure and allows flows to be compared according to the similarity of their autocorrelation profiles.

We then compute pairwise dissimilarities between all $N^2$ ACF feature vectors using the Euclidean distance:

\begin{equation}
\mathbf{D}^{(\mathrm{acf})} \in \mathbb{R}^{N^2 \times N^2},
\qquad
\mathbf{D}_{ij}^{(\mathrm{acf})}
=
\left\|
\mathbf{f}_i^{(\mathrm{acf})}
-
\mathbf{f}_j^{(\mathrm{acf})}
\right\|_2.
\end{equation}

For the ACF representation, we use average linkage in the HAC stage. Under average linkage, the distance between two clusters $U$ and $V$ is defined as the average pairwise distance between their members~\cite{dasgupta2005performance}:

\begin{equation}\label{eq:average_linkage}
d_{\mathrm{AL}}(U,V)
=
\frac{1}{|U||V|}
\sum_{x_i \in U}
\sum_{x_j \in V}
\mathbf{D}_{ij}^{(\mathrm{acf})},
\end{equation}
where $|U|$ and $|V|$ denote the cardinalities of clusters $U$ and $V$, respectively. 

Thus, the ACF-based representation clusters traffic flows according to similarities in their temporal correlation structure rather than their raw traffic values or node locations. These clusters provide a temporal-dependence-based grouping of flows for the cluster-specific prediction models evaluated in this work.

\subsection{PSD Representation}

For the PSD-based representation, each traffic flow $x_i \in \mathbb{R}^T$ is characterized in the frequency domain by its power spectral density (PSD). Whereas the ACF representation emphasizes lag-dependent temporal dependence, the PSD representation highlights the distribution of signal power across frequencies and is therefore well-suited for identifying flows with similar periodic behavior.

Let $S_i(f)$ denote the PSD of flow $x_i$ at frequency $f$. In practice, the PSD is estimated over a discrete set of frequencies $\{f_1,f_2,\dots,f_L\}$, yielding the feature vector
\begin{equation}
\mathbf{f}_i^{(\mathrm{psd})}
=
\begin{bmatrix}
S_i(f_1) & S_i(f_2) & \cdots & S_i(f_L)
\end{bmatrix}^{\top}.
\end{equation}
This representation captures the spectral content of the flow and enables grouping of time series with similar dominant frequencies, even when their raw time-domain behavior differs. Prior to clustering, the PSD features may be normalized to reduce the influence of scale differences across flows. We compute PSD using Welch's method~\cite{Welch1967}. 

We then compute pairwise dissimilarities between all $N^2$ PSD feature vectors using the Euclidean distance:
\begin{equation}
\mathbf{D}^{(\mathrm{psd})} \in \mathbb{R}^{N^2 \times N^2},
\qquad
\mathbf{D}_{ij}^{(\mathrm{psd})}
=
\left\|
\mathbf{f}_i^{(\mathrm{psd})}
-
\mathbf{f}_j^{(\mathrm{psd})}
\right\|_2.
\end{equation}

For the PSD representation, we use average linkage in the HAC stage as described by~\eqref{eq:average_linkage}. Thus, the PSD-based representation complements the ACF representation by clustering traffic flows according to similarities in how their signal power is distributed across frequencies, rather than their raw time-domain values. These clusters provide a spectral-behavior-based grouping of flows for the cluster-specific prediction models evaluated in this work.

\subsection{Na\"{i}ve Clustering Baseline}

Finally, to assess whether performance gains arise from representation-driven clustering or simply from partitioning the prediction task into smaller subproblems, we propose a na\"{i}ve clustering baseline. Under this method, the $N^2$ traffic flows are assigned randomly to $K$ clusters, without using any feature representation, distance metric, or similarity measure.

Formally, let $\mathcal{X}=\{x_1,x_2,\dots,x_{N^2}\}$ denote the set of traffic-flow time series. The na\"{i}ve baseline defines a cluster assignment map
\begin{equation}
c : \mathcal{X} \rightarrow \{1,2,\dots,K\},
\end{equation}
where each flow $x_i$ is assigned to a cluster label $c(x_i)$ according to a random assignment procedure. This produces a partition of the flow set into $K$ disjoint clusters,
\begin{equation}
\mathcal{C} = \{C_1,C_2,\dots,C_K\},
\end{equation}
such that
\begin{equation}
\bigcup_{k=1}^{K} C_k = \mathcal{X}, \qquad C_k \cap C_{\ell} = \emptyset \;\; \text{for } k \neq \ell.
\end{equation}

Unlike the histogram, ACF, and PSD-based approaches, na\"{i}ve clustering does not attempt to group flows according to shared distributional, temporal, or spectral structure. Instead, it serves as a control condition that isolates the effect of decomposing the overall TM prediction task into smaller cluster-specific prediction problems. Comparing this baseline with representation-based clustering therefore helps determine whether observed improvements stem from meaningful structure in the induced clusters or from the act of partitioning alone.

\section{Experimental Settings}\label{sec:experimental_settings}
\label{sec:experiemental_settings}
This section describes the datasets, preprocessing steps, benchmark models, evaluation metrics, and implementation details used in our experiments.

\subsection{Datasets \& Preprocessing}
We evaluate our prediction methods using real traffic matrix data from the Abilene~\cite{AbileneDataset} and G\'{E}ANT~\cite{GEANTDataset} networks. The Abilene network has 12 nodes and 30 links, with TMs collected at 5-minute intervals over 24 weeks. The G\'{E}ANT network has 23 nodes and 74 links, with TMs collected at 15-minute intervals over 4 months. We apply an 80/20 train-test split, and 10\% of the training samples are used as a validation set.

Training samples are constructed using a sliding window of length $L$, where the first $L-1$ traffic matrices serve as inputs and the $L$th matrix is used as the prediction target~\cite{azzouni2018neutm}. In our experiments, we use 10 historical TMs as input for prediction. We apply min-max normalization to scale traffic flows to the range $[0,1]$ during training.

\subsection{Benchmark Models}
We benchmark the clustering-based approaches against three state-of-the-art EM models: Prophet~\cite{zhang2023prophet}, ARCNN~\cite{gao2020incorporating}, and a standalone GRU~\cite{azzouni2018neutm}. Prophet is a traffic-engineering-centric framework that uses a GRU model with an angle-based loss function for TM prediction. ARCNN is a spatial-temporal model that combines CNNs to capture inter-flow correlations, RNNs to model intra-flow dependencies~\cite{ramakrishnan2018network}, and an attention mechanism for long-range temporal modeling. The standalone GRU baseline represents a simpler entire-matrix predictor. All three baselines operate under the EM prediction perspective. We also compare against the local prediction perspective by training $K=N^2$ GRU models, one per traffic flow~\cite{zheng2022flow}. 

\subsection{Evaluation Metrics}
We evaluate prediction performance using root mean squared error (RMSE):
\begin{equation}
\mathrm{RMSE} = \sqrt{\frac{1}{n}\sum_{i=1}^{n}(\mathrm{TM}_i-\mathrm{\hat{TM}}_i)^2}
\end{equation}

\noindent where $n$ denotes the number of test samples, $\mathrm{TM}_i$ denotes the ground-truth traffic matrix, and $\mathrm{\hat{TM}}_i$ denotes the predicted traffic matrix. We report RMSE for both normalized and non-normalized testing sets.

To compare clustering assignments across methods, we use the Adjusted Rand Index (ARI) and Normalized Mutual Information (NMI). ARI measures pairwise agreement between two clusterings while correcting for chance, whereas NMI measures the shared information between two clusterings. ARI is useful for assessing consistency in pairwise assignments, while NMI is better suited for comparing methods that may produce different levels of cluster granularity. We refer interested readers to~\cite{hubert1985comparing} and~\cite{strehl2002cluster} for the full definitions. We keep the number of clusters $K$ consistent across methods when computing ARI and NMI.

\subsection{Implementation Details}
All DL models are implemented using PyTorch. For all clustering approaches and benchmark models, excluding ARCNN, we use a one-layer GRU with a hidden size of 200, following configurations used in prior work~\cite{liu2019traffic,zhang2023prophet,cash2025improving}. Models are trained for 100 epochs with a batch size of 32 using the Adam optimizer~\cite{kingma2014adam} with a learning rate of 0.001. Early stopping is applied with a patience of 5 and a minimum delta of $1\times10^{-5}$ to prevent overfitting. These parameter choices are consistent with settings commonly adopted in prior TM prediction studies~\cite{valadarsky2017learning,ramakrishnan2018network}.

For the histogram representation, we use 50 bins for both datasets. For the ACF representation, we select lags $\ell$ according to the sampling interval of each dataset so as to capture short-, medium-, and long-range temporal structure. Specifically, short-range lags cover up to two hours, medium-range lags correspond to approximately three to six hours, and long-range lags include half-day and full-day samples. For the PSD representation, the sampling frequency $f_s$ is chosen to match the number of samples per hour: $f_s=12$ for Abilene and $f_s=4$ for G\'{E}ANT.

\section{Experimental Results}
\label{sec:results}
This section presents the experimental results for the proposed clustering-based framework and the baseline TM prediction methods. In addition to overall prediction performance, we investigate the effect of the number of clusters, $K$ on performance for each representation. We further analyze the cluster structures produced by each method using ARI, NMI, and cluster size distributions. 

\subsection{Impact of $K$-Sweep}
\begin{figure*}[!t]
    \centering
    \begin{subfigure}{\textwidth}
        \centering
        \includegraphics[width=\textwidth]{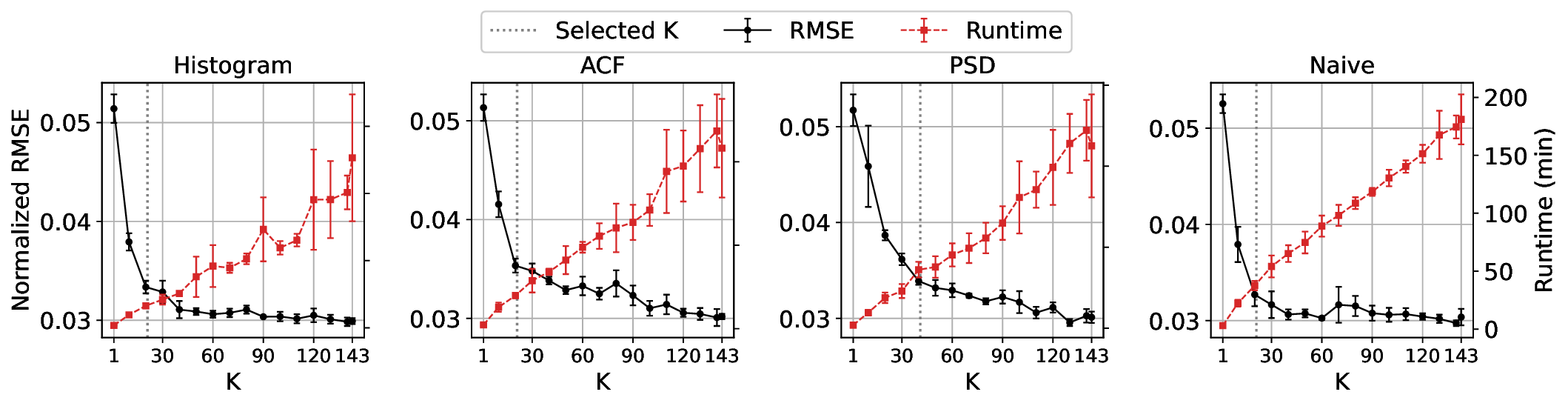}
        \caption{Abilene}
    \end{subfigure}
    \vspace{1em}
    \begin{subfigure}{\textwidth}
        \centering
        \includegraphics[width=\textwidth]{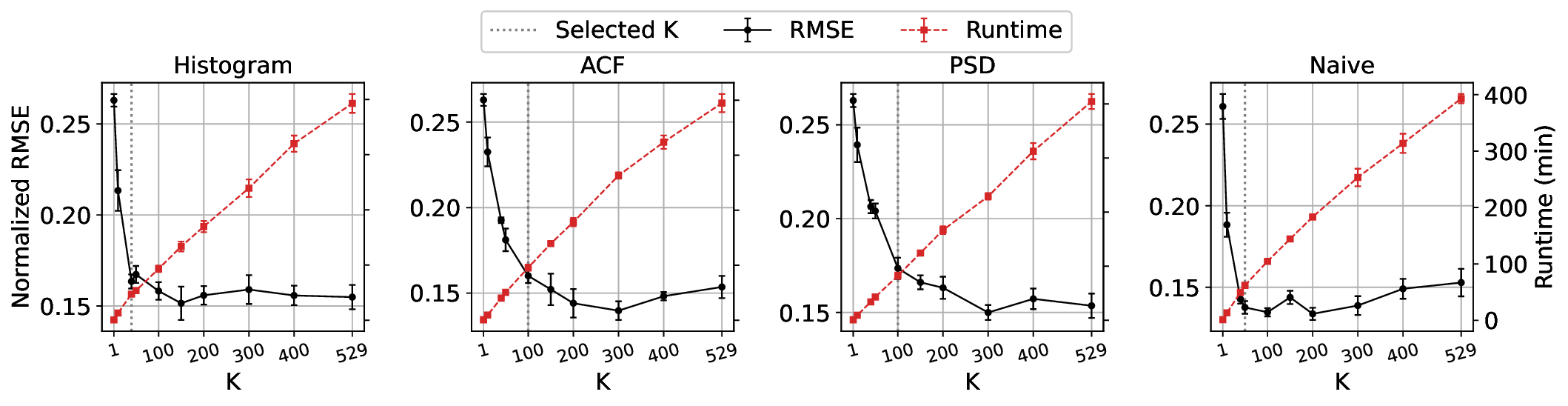}
        \caption{G\'EANT}
    \end{subfigure}
    \caption{$K$-sweep results for each traffic-flow representation. Each panel shows normalized RMSE and runtime as functions of the number of clusters, \(K\). The dotted vertical line indicates the selected value of \(K\) used in subsequent experiments. Specific $K$ values are detailed in Table~\ref{tab:optimal-k}. }
    \label{fig:k_sweep_all}
\end{figure*}
\begin{table}[!tb]
\caption{Selected values of $K$ and runtime in minutes obtained from the knee point of the RMSE-versus-$K$ curves for each clustering method.}
\label{tab:optimal-k}
\begin{center}
\small
\resizebox{\columnwidth}{!}{
\begin{tabular}{c c c c c c c c c}
\hline
\textbf{Network} 
& \multicolumn{2}{c}{\textbf{Histogram}} 
& \multicolumn{2}{c}{\textbf{ACF}} 
& \multicolumn{2}{c}{\textbf{PSD}} 
& \multicolumn{2}{c}{\textbf{Na\"{i}ve}} \\
\cline{2-9}
& $K$ & Runtime 
& $K$ & Runtime 
& $K$ & Runtime 
& $K$ & Runtime \\
\hline
Abilene 
& 21 & 37
& 21 &  36
& 41 & 55
& 21 & 44 \\

G\'EANT 
& 40 & 46
& 100 & 89
& 100 & 91 
& 50 & 67 \\
\hline
\end{tabular}
}
\end{center}
\end{table}

We first examine how the number of clusters $K$ affects prediction performance and runtime, as shown in Fig.~\ref{fig:k_sweep_all}. For each representation and dataset, we sweep $K$ from $1$ to $N^2$, corresponding to the range between entire-matrix prediction ($K=1$) and local prediction ($K=N^2$). This corresponds to $K \in [1,144]$ for Abilene and $K \in [1,529]$ for G\'EANT. For computational efficiency, we increment $K$ by $\Delta=10$ and repeat each sweep five times. Fig.~\ref{fig:k_sweep_all} reports the mean normalized RMSE and runtime across trials, with error bars indicating variability.

Across all methods and both datasets, RMSE decreases sharply as $K$ increases from small values and then begins to level off, while runtime increases steadily with larger $K$. This behavior indicates a diminishing return regime in which further increasing the number of clusters yields only modest gains in RMSE at substantially higher computational cost.

To select the operating point for each clustering method, we apply the Kneedle algorithm~\cite{satopaa2011finding} to the RMSE-versus-$K$ curves and identify the knee point as the optimal value of $K$. The selected $K$ values are summarized in Table~\ref{tab:optimal-k}, along with the corresponding runtimes reported in minutes. In general, the resulting operating points correspond to a relatively small fraction of the full set of traffic flows: approximately $14$--$30\%$ of the $144$ flows for Abilene and $7$--$19\%$ of the $529$ flows for G\'EANT. 

These operating points also provide a favorable tradeoff between runtime and prediction accuracy. For the Abilene dataset, predicting every traffic flow individually requires approximately $120$--$150$ minutes, whereas the clustering-based approaches require roughly one quarter of that time. Similarly, for the G\'EANT dataset, local prediction requires approximately $400$ minutes, while the clustering-based approaches again reduce runtime substantially. These results suggest that substantial prediction gains can be achieved well before reaching the local prediction extreme.

A notable observation is that the na\"{i}ve clustering baseline also yields lower RMSE than the $K=1$ entire-matrix setting on both datasets. This suggests that a substantial portion of the improvement comes from decomposing the prediction task into smaller subproblems, rather than exclusively from the specific representation used to form the clusters. Near the selected knee-point operating values of $K$, the representation appears to play a secondary role, as the clustering methods yield broadly comparable prediction performance. However, when the number of clusters is constrained below these operating points, the choice of representation may become more important because it determines which flows are grouped together under a limited model budget.

\subsection{Overall Prediction Performance}
\begin{table*}[!tb]
\caption{Test RMSE in Mbps for clustering using optimal $K$ and benchmark prediction methods on the Abilene and G\'EANT datasets.}
\label{tab:combined-error}
\centering
\resizebox{0.8\textwidth}{!}{
\begin{tabular}{ccccccccc}
\hline
\textbf{Dataset} & \textbf{Histogram} & \textbf{ACF} & \textbf{PSD} & \textbf{Naive} & \textbf{Prophet} & \textbf{ARCNN} & \textbf{EM} & \textbf{Local} \\
\hline
Abilene & 0.64 & 0.755 & 0.597 & 0.688 & 3.35 & 2.50 & 1.28 & 0.189 \\
G\'EANT & 0.014 & 0.027 & 0.014 & 0.014 & 0.067 & 0.041 & 0.040 & 0.003 \\

\end{tabular}
}
\end{table*}

Next, we compare the overall prediction performance of the clustering methods against the benchmark models. Table~\ref{tab:combined-error} reports the test RMSE in Mbps for both the clustering methods and the baselines. In this setting, we compare RMSE for $K$ determined from Table~\ref{tab:optimal-k}. Across both datasets, clustering consistently reduces RMSE relative to the entire-matrix benchmark models, while local prediction achieves the lowest error overall. For the Abilene dataset, PSD clustering provides the strongest performance among the clustering approaches, with an RMSE of 0.597 Mbps. For G\'EANT, the histogram-, PSD-, and na\"{i}ve-clustering methods achieve the lowest RMSE among the clustering methods, all with an error of 0.014 Mbps. These results indicate that clustering offers an effective middle ground: it improves prediction accuracy relative to global entire-matrix forecasting while avoiding the full computational cost of the local prediction regime.

\begin{figure}
\begin{subfigure}{0.47\textwidth}
    \centering
    \includegraphics[width=\textwidth]{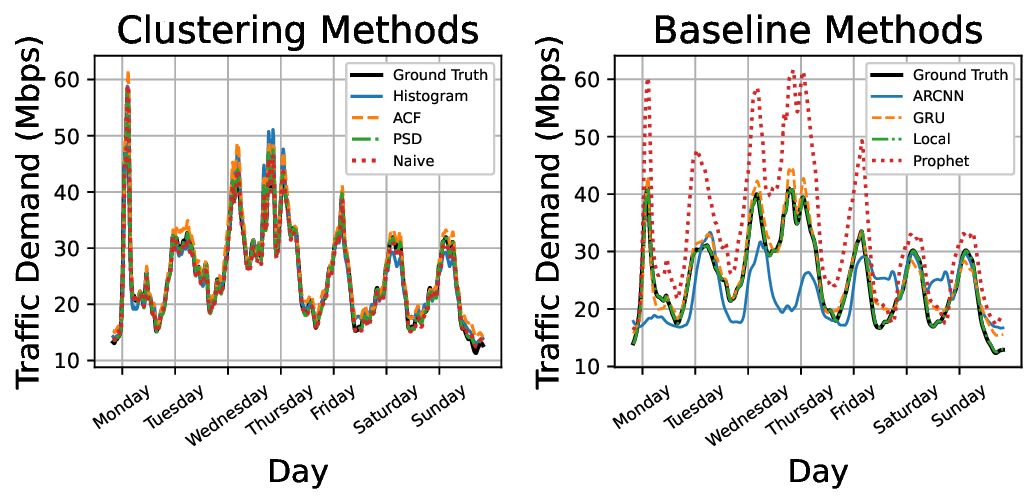}
    \caption{Abilene}
    \label{fig:preds_abilene}
\end{subfigure}
\hfill
\begin{subfigure}{0.47\textwidth}
    \centering
    \includegraphics[width=\textwidth]{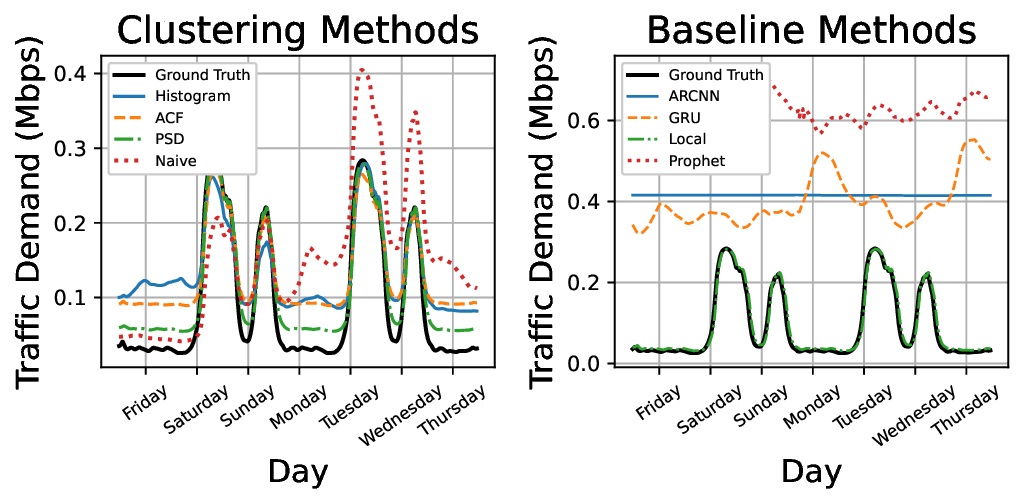}
    \caption{G\'EANT}
    \label{fig:preds_geant}
\end{subfigure}
\caption{Representative traffic-flow predictions for the Abilene and G\'EANT datasets. The clustering-based methods more closely track the ground truth than the entire-matrix baselines while remaining less computationally expensive than local prediction.}
\label{fig:overall_prediction}
\end{figure}

Fig.~\ref{fig:overall_prediction} provides a visual comparison of representative traffic-flow predictions for both the Abilene and G\'EANT datasets. For Abilene, the clustering-based methods more closely follow the ground-truth trajectory than the benchmark entire-matrix methods, which often overestimate or underestimate the target flow. In particular, the more complex benchmark models, namely, ARCNN and Prophet, do not track the ground truth as effectively as the clustering-based approaches, showing that greater architectural complexity in the entire-matrix setting does not necessarily yield better prediction performance.

For G\'EANT, prediction is more challenging overall, and all benchmark methods exhibit greater difficulty tracking the ground truth. ARCNN, in particular, frequently produces low-variance predictions that fail to capture the bursty dynamics of the representative flow, suggesting that this architecture does not scale as effectively to the larger and more heterogeneous traffic matrix. Although the clustering-based predictions do not track the ground truth as closely as in the Abilene case, most of them still capture the major peaks of the representative flow. Among these, PSD clustering provides the closest visual agreement with the G\'EANT ground truth.

\subsection{Analysis of Clustering Structure}
\begin{table}[!tb]
\caption{Pairwise clustering similarity measured by NMI and ARI for each dataset.}
\label{tab:pairwise_nmi_ari}
\centering
\begin{tabular}{c c c c}
\hline
\textbf{Dataset} & \textbf{Comparison} & \textbf{NMI} & \textbf{ARI} \\
\hline
\multirow{6}{*}{Abilene}
& Hist. vs ACF   & 0.47 & 0.091  \\
& Hist. vs PSD   & 0.44 & 0.098  \\
& Hist. vs Naive & 0.40 & 0.016 \\
& ACF vs PSD     & 0.56 & 0.24  \\
& ACF vs Naive   & 0.34 & -0.008 \\
& PSD vs Naive   & 0.32 & 0.0 \\
\hline
\multirow{6}{*}{G\'EANT}
& Hist. vs ACF   & 0.27     & 0.013      \\
& Hist. vs PSD   & 0.25     & 0.032      \\
& Hist. vs Naive & 0.28     & 0.001      \\
& ACF vs PSD     & 0.29     & 0.06      \\
& ACF vs Naive   & 0.24    & 0.002      \\
& PSD vs Naive   & 0.19     & 0.002      \\
\hline
\end{tabular}
\end{table}
To better understand why different clustering methods often yield similar prediction performance, we analyze the structure of the induced clusters from three perspectives: pairwise clustering similarity, cluster size statistics, and consistency of prediction errors across methods. For the NMI and ARI analysis, we fix the number of clusters to be the same across methods in order to enable a fair structural comparison. Specifically, we use $K=21$ for Abilene and $K=40$ for G\'EANT.

Table~\ref{tab:pairwise_nmi_ari} summarizes the pairwise clustering similarity measured by ARI and NMI. Across most method pairs and both datasets, the ARI values are low, indicating weak agreement in the exact flow-to-cluster assignments beyond chance. Thus, even when the same number of clusters $K$ is used, different representations generally partition the traffic flows into different groups. The clearest exception occurs for the ACF and PSD representations on Abilene, where the ARI reaches 0.24, suggesting a modest level of agreement in pairwise assignments. This is consistent with the fact that both representations capture temporal dependence and periodic behavior, albeit in different forms.

A different pattern emerges in the NMI results. While the ARI values indicate weak agreement in exact flow-to-cluster assignments, the NMI values suggest that some amount of information is shared across partitions. However, because the na\"{i}ve clustering baseline also exhibits comparable NMI values in several cases, especially for G\'EANT, these values should not be interpreted as strong evidence that the representation-based methods recover a common underlying cluster structure. Instead, the NMI results likely reflect a baseline level of agreement that can arise from random partitioning, the fixed number of clusters, and the fact that all methods are applied to the same set of traffic flow time series. Therefore, although the representations induce different cluster assignments, the NMI results should be viewed as limited evidence of broad structural overlap rather than definitive evidence of a shared clustering structure.

Taken together with the RMSE results, this suggests that comparable prediction performance does not require the clustering methods to recover the same partition of traffic flows. Rather, the benefit of clustering appears to come primarily from decomposing the prediction task into smaller subproblems, with the particular partition playing a more secondary role at the selected operating values of $K$.

Next, we examine the cluster size statistics reported in Table~\ref{tab:cluster_size_stats}, including the minimum, mean, and maximum number of traffic flows per cluster, along with the number and percentage of singleton clusters. These statistics characterize how balanced or fragmented each partition is and therefore help clarify whether differences in prediction performance are associated with differences in cluster granularity.

Several trends emerge. First, na\"{i}ve clustering produces the most balanced partitions on both datasets, with no singleton clusters and very narrow ranges between the minimum and maximum cluster sizes. In contrast, the representation-based methods often yield more uneven partitions. This effect is especially pronounced for PSD, which produces a large number of singleton clusters, particularly on G\'EANT, where 75\% of the clusters contain only a single flow. ACF also generates substantial fragmentation, while histogram-based clustering tends to produce fewer singleton clusters and somewhat larger dominant groups. These results indicate that the representations induce meaningfully different partition structures, ranging from relatively balanced groupings to highly fragmented decompositions. Importantly, these structural differences do not translate into equally large differences in RMSE, further suggesting that prediction gains may depend more on decomposing the task than on recovering any single preferred cluster-size distribution.

\begin{table}[!tb]
\caption{RMSE error consistency across representations}
\label{tab:error_correlation}
\centering
\begin{tabular}{ccc}
\hline
\textbf{Dataset} & \textbf{Comparison} & \textbf{Correlation of Prediction Error} \\
\hline
\multirow{6}{*}{Abilene}
& Hist. vs ACF      & 0.88 \\
& Hist. vs PSD      & 0.89 \\
& Hist. vs Na\"{i}ve & 0.91 \\
& ACF vs PSD        & 0.87 \\
& ACF vs Na\"{i}ve   & 0.84 \\
& PSD vs Na\"{i}ve   & 0.85 \\
\hline
\multirow{6}{*}{G\'EANT}
& Hist. vs ACF      & 0.87 \\
& Hist. vs PSD      & 0.83 \\
& Hist. vs Na\"{i}ve & 0.68 \\
& ACF vs PSD        & 0.88 \\
& ACF vs Na\"{i}ve   & 0.78 \\
& PSD vs Na\"{i}ve   & 0.56 \\
\hline
\end{tabular}
\end{table}

Finally, Table~\ref{tab:error_correlation} reports the Pearson correlation between the per-flow prediction error vectors produced by different clustering methods. This analysis is intended to determine whether methods with different cluster structures tend to make errors on the same traffic flows. For each method, we form a vector containing the prediction error for every flow, yielding vectors of length 144 and 529 for Abilene and G\'EANT, respectively, and then compute pairwise Pearson correlations between these vectors.

The resulting correlations are high across nearly all method pairs, particularly for Abilene, where all values exceed $0.84$. This indicates that even when the cluster assignments differ, the same traffic flows tend to remain easy or difficult to predict across methods. A similar trend appears on G\'EANT, although the correlations are somewhat lower for comparisons involving na\"{i}ve clustering, especially PSD versus na\"{i}ve. This suggests that while clustering structure can alter the partition of flows, it does not fully change which flows are intrinsically challenging to predict. In other words, some of the prediction difficulty appears to arise from properties of the flows themselves rather than from the particular clustering representation used.

\begin{table}[!t]
\caption{Cluster size statistics at the selected value of $K$. Cluster size is measured as the number of traffic flows assigned to each cluster.}
\label{tab:cluster_size_stats}
\centering
\resizebox{\columnwidth}{!}{
\begin{tabular}{cccccccc}
\hline
\textbf{Dataset} & \textbf{Method} & \textbf{$K$} & \textbf{Min} & \textbf{Mean} & \textbf{Max} & \textbf{\# Singletons} & \textbf{Singleton \%} \\
\hline
\multirow{4}{*}{Abilene}  
& Histogram & 21 & 1 & 6.86 & 20 & 2 & 9.52 \\
& ACF & 21 & 1 & 6.86 & 31 & 6 & 28.57 \\
& PSD & 41 & 1 & 3.51 & 17 & 19 & 46.34 \\
& Naive & 21 & 6 & 6.86 & 7 & 0 & 0.00 \\
\hline
\multirow{4}{*}{G\'EANT}
& Histogram & 40 & 1 & 13.22 & 79 & 5 & 12.50 \\
& ACF & 100 & 1 & 5.29 & 55 & 37 & 37.00 \\
& PSD & 100 & 1 & 5.29 & 76 & 75 & 75.00 \\
& Naive & 50 & 10 & 10.58 & 11 & 0 & 0.00 \\

\end{tabular}
}
\end{table}

\section{Discussion}
\label{sec:discussion}
This section discusses the broader implications of the experimental results and synthesizes the main findings across the preceding subsections. Overall, the results show that clustering provides a meaningful improvement over EM prediction methods while remaining substantially less costly than local prediction. At the same time, the results reveal that the benefit of clustering does not appear to depend on recovering one unique grouping of traffic flows. Instead, the primary gain appears to come from decomposing the TM prediction task into smaller subproblems. Near the selected operating values of $K$, the exact clustering representation plays a more limited role in determining aggregate RMSE, although it may still influence performance when the number of available clusters is constrained.

This interpretation is most clearly reflected in the comparison between EM, clustered, and local prediction reported in Table~\ref{tab:combined-error}. Local prediction achieves the lowest RMSE, but it does so at the cost of substantially higher runtime because it requires training one model per traffic flow. On average, local prediction requires approximately $150$--$200$ minutes for the Abilene dataset and about $400$ minutes for the G\'EANT dataset. In this sense, clustering offers a practical middle ground between predictive accuracy and computational efficiency. These results suggest that full local prediction is not required to obtain most of the achievable improvement in RMSE.

This conclusion is further supported by the $K$-sweep results. Moving away from the EM setting of $K=1$ yields rapid gains in prediction performance, after which the RMSE curves exhibit diminishing returns. Notably, the point of diminishing returns occurs at values of $K$ that are much smaller than the total number of flows. Thus, most of the prediction benefit can be achieved at moderate cluster counts, well before reaching the local-prediction regime. However, the choice of representation may be more important when operating below these selected values of $K$, since the representation determines how effectively a limited number of clusters captures similarities among traffic flows.

Interestingly, the clustering methods produce broadly similar RMSE despite inducing different partitions of the traffic flows. The ARI values show that exact flow assignments differ substantially across representations, while the NMI results indicate a limited degree of shared information across partitions, some of which may reflect baseline agreement rather than a common underlying cluster structure. In other words, representation strongly affects how flows are grouped, but near the selected operating points these structural differences do not translate into equally large differences in prediction error. This helps explain why histogram, ACF, PSD, and na\"{i}ve clustering can produce different clusterings while still achieving comparable overall RMSE.

This interpretation is reinforced by the na\"{i}ve clustering baseline. Since na\"{i}ve clustering also improves over EM prediction, part of the performance gain must come simply from partitioning the prediction task into smaller subproblems. Thus, some of the benefit of clustering appears to arise from decomposition itself, rather than solely from discovering semantically meaningful groupings of traffic flows. However, this does not imply that the representation is irrelevant; rather, the results suggest that representation is less critical once the number of clusters is sufficiently large, but may play a larger role when the prediction framework is restricted to fewer clusters.

Finally, the results show that dataset complexity and network scale affect both prediction difficulty and the relative effectiveness of clustering. Abilene, as the smaller network, exhibits stronger agreement across methods and easier tracking of representative flows. In contrast, the G\'EANT dataset is more heterogeneous and harder to predict. This is particularly evident in the ARCNN baseline, which frequently produces low-variance predictions on G\'EANT. In this sense, the G\'EANT results expose the limitations of complex global architectures and further highlight the practical value of clustered prediction.

\section{Conclusion}
\label{sec:conclusion}
In this paper, we propose clustering-based TM prediction as an alternative to EM and local prediction. We evaluated histogram, ACF, PSD, and na\"{i}ve clustering strategies within a common prediction framework on the Abilene and G\'EANT datasets. The results showed that clustering consistently improves over EM baselines while remaining substantially less costly in terms of run time than local prediction. 

The experiments further showed that strong prediction performance can be achieved at moderate values of $K$, well before reaching the local-prediction regime. Additionally, the clustering-based prediction models require roughly one quarter of the runtime of local prediction. Although the clustering methods often produced different flow partitions, they yielded broadly similar RMSE values, indicating that clustering can improve TM prediction even when the exact grouping structure varies across representations.

Overall, these findings support clustering as a practical and effective framework for TM prediction, particularly when balancing predictive accuracy against computational runtime. More broadly, the results suggest that decomposing the prediction task into smaller subproblems is more beneficial than a specific representation. Therefore, when the objective is to obtain the benefits of clustering with minimal implementation overhead, we recommend na\"{i}ve clustering as a simple default strategy, since it does not require additional effort for transforming traffic flows into histogram, ACF, or PSD representations and achieves comparable RMSE in our experiments.

Future work will extend this evaluation to additional network datasets and traffic conditions to better characterize when clustering-based TM prediction is most effective. The similarity in RMSE across clustering methods also suggests the need to study flow-level predictability directly, particularly to identify which traffic flows benefit most from more expressive prediction models.

\printbibliography

\end{document}